\documentclass[12pt,a4paper]{article}
\usepackage{graphicx}
\usepackage[cp1251]{inputenc}
\usepackage{amssymb, amsfonts, amsmath}
\pdfoutput=1
\numberwithin{equation}{section}
\date{}
\usepackage{amscd}
\begin{document}

\title{Algebra of the symmetry operators of the Klein-Gordon-Fock equation for the case when groups of motions $G_3$ act transitively on null subsurfaces of spacetime}

 \author{V.V. Obukhov}

 \maketitle
 
  Institute of Scietific Research and Development,

 Tomsk State Pedagogical University (TSPU), 634041, Tomsk, Russia;

 \quad

Lab. for Theor. Cosmology, International Centre of Gravity and Cosmos,

Tomsk State University of Control Systems and Radioelectronics (TUSUR),

 634050 Tomsk, Russia.
 
 \quad


            and tensors, integrals of motion.

            \quad

\section{Introduction}

The Klein-Gordon-Fock equation describes the dynamics of massive spinless test particles interacting with fields of a gauge nature. It is used to study quantum field effects in external electromagnetic and gravitational fields for scalar particles and to build approximate models for fermions. In this case, the problem of finding the exact basic solutions of the Klein-Gordon-Fock equation in the external intensive fields is of great importance. The basic solution is a common eigenfunction of the complete set of symmetry operators. In order to obtain the exact basic solution, it is necessary to find a commutative algebra consisting of three linear differential symmetry operators no more than quadratic in momenta. The problem of constructing such algebras has been sufficiently studied. Suppose this algebra forms the traditional complete set of symmetry operators for the Klein-Gordon-Fock equation. In this case, spacetime admits a complete set of geometric objects, consisting of mutually commuting vector and tensor Killing fields, and belongs to the set of Stackel spaces.

A Stackel space  $V_n$  is an $n$-dimensional Riemannian space of arbitrary signature, in which the free $n$-dimensional Hamilton--Jacobi equation for a massive test particle is integrated by the method of complete separation of variables. The Stackel space admits the complete set of Killing fields. It is proved that the $n$-dimensional Klein-Gordon-Fock equation can be integrated by the method of complete separation of variables only if it admits the traditional complete set of symmetry operators.

This is possible only in certain classes of Stackel spaces. The method for finding basic solutions based on the complete separation of variables is also called the commutative integration method. For information on the method of complete separation of variables and the results obtained with its help, see \cite{VSh1} -\cite{A}, and the articles cited there.

In the paper \cite{ASh1}--\cite{ASh2} a method for the integration of linear partial differential equations in $n$-dimensional Riemannian spaces (and also in the Hamilton--Jacobi equation) of arbitrary signature admitting noncommutative groups of motions  $G_r$  was proposed (these spaces are also denoted by $V_n$. The algebras of the symmetry operators of the Klein-Gordon-Fock equation of rank \quad $r \quad (n-1 \leq r\leq n$), constructed using the algebras of operators of the noncommutative group of motions of the space  $V_n$, are complemented to a commutative algebra by the operators of differentiation of the first order in $n$ essential parameters. Basic solutions are found using these parameters. By analogy with the method of complete separation of variables, such algebras are called complete sets, and the integration method is called noncommutative integration.  The methods are related because they both reduce the problem of finding the basic solution of the test particle equation of motion to the problem of integrating systems of ordinary differential equations.

The noncommutative integration method is based on the complete classification of spacetime manifolds admitting groups of motions described in the book \cite{Petrov}. The method made it possible to considerably extend the set of fields in which the construction of a complete system of solutions of the classical and quantum equations of a charged test particle motion is reduced to the integration of compatible systems of first-order differential equations.
For further development of the method and its application in gravitational theory, it is necessary (using the proposed classification) to make a classification of electromagnetic fields in which the classical and quantum equations of motion of a charged test particle (the Hamilton--Jacobi and Klein-Gordon-Fock equations) admit noncommutative algebras of symmetry operators that are linear in momenta. Such electromagnetic fields are called admissible.

For the first time, this problem was formulated and partially solved in \cite{Mag1}--\cite{Mag2}, where the potentials of all admissible electromagnetic fields in spacetime manifolds, admitting the transitive action of four-parameter groups of motions, are given. A similar classification problem was solved for homogeneous spaces with a three-parameter group of motions (\cite{OVV1}) and for spaces with a two-parameter movement group (\cite{OVV2}). Moreover, the problem is solved for the case when a four-parameter groups of motions with a three-dimensional hypersurface of transitivity acts on a  spacetime manifold (\cite{OVV3}). In the present work, the classification of admissible fields is carried out when the three-parameter groups of motions $G_3$  act transitively on isotropic hypersurfaces of the space  $V_4$   with a spacetime signature. We have found all relevant admissible electromagnetic fields.

The article is organized as follows.

The second section contains the necessary information and definitions required for the implementation of this classification. The conditions that must be met by admissible electromagnetic fields are obtained and investigated for compatibility.

In the third and fifth sections, the obtained conditions are used to find the potentials of the admissible electromagnetic field for resolvable groups of motion. The cases of groups with a singular operator are considered separately.

In the fourth section, unsolvable groups of motion are considered.

In conclusion, possible applications of the obtained results are considered.

\section{Admissible electromagnetic fields}

\subsection{Conditions for the existence of the symmetry operators algebra in the case of a charged test particle motion}

\quad

Consider a  spacetime manifold $V_4$ on an null hypersurface on which the three-parameter movement groups $G_3$ acts transitively. The coordinate indices of the canonical coordinate system\quad $[u^i] $ of the space $V_4$ are denoted by lower case Latin letters:\quad $i, j, k = 0, 1 \dots 3$.\quad The coordinate indices of the canonical coordinate system on the isotropic hypersurfaces $V^*_3$ will be denoted by lower case Greek letters:\quad $\alpha, \beta, \gamma $=1, \dots 3.\quad A non-ignored variable is indexed $0$. The repeated superscripts and subscripts are summed within the limits of the indices change. The papers \cite{Mag1}, \cite{OVV1} show that for a charged test particle in an external electromagnetic field with potential $A_i$ the Hamilton-Jacobi, and Klein-Gordon-Fock equations:

\begin{equation}\label{1}
H=g^{ij}P_iP_j=m, \quad P_i=p_i+A_i,\quad p_i=\partial_i\varphi
\end{equation}
\begin{equation}\label{9}
\hat{H}\varphi=(g^{ij}\hat{P}_i\hat{P}_j)\varphi = m\varphi, \quad \hat{P}_j = -\imath \hat{\nabla}_i + A_i,
\end{equation}
admit algebras of symmetry operators (in the case of the Hamilton--Jacobi equation, integrals of motion) in the same electromagnetic fields.

Here \quad $\hat{\nabla}_i$ \quad  is the operator of the covariant derivative corresponding to the operator of the partial derivative \quad $\hat{\partial}_i =\imath \hat{p}_i$ \quad along the coordinate $u^i, \varphi$ is the field of a scalar particle with mass $m$.

Therefore, for the implementation of the admissible electromagnetic fields classification, the Hamilton--Jacobi equation would suffice. The integrals of motion of the free Hamilton-Jacobi equation have the form:
\begin{equation}\label{2}
Y_\alpha=\xi_\alpha^i p_i,
\end{equation}
where $\xi^{j}_\alpha$ are the Killing vector fields satisfying the equations:

\begin{equation}\label{3}
    g^{ik}{\xi^{j}_\alpha}_{,k}+g^{jk}{\xi^{i}_\alpha}_{,k}-g^{ij}_{,k}\xi_\alpha^k=0.
\end{equation}
$\xi^{j}_\alpha$ defines the movement groups $G_3$ of the space $V_4$. It can be shown that if the equation \eqref{1} has  $r$   independent integrals of motion of the first order,  then these operators have the form \eqref{2}. the Hamilton-Jacobi equation \eqref{1} admits a motion integral of the form if $H$  and  $Y_\alpha$ commute with respect to the Poisson brackets:
\begin{equation}\label{4}
[H,Y_\alpha]_P=\frac{\partial H}{\partial p_i}\frac{\partial Y_\alpha}{\partial x^i} - \frac{\partial H}{\partial x^i}\frac{\partial Y_\alpha}{\partial p_i}=
\end{equation}
$$
(g^{ik}{\zeta^{j}_\alpha}_{,k}+g^{jk}{\zeta^{i}_\alpha}_{,k}-g^{ij}_{,k}\zeta_\alpha^k)P_i P_j + 2g^{ik}(\xi^{j}_\alpha F_{ji}+(\zeta_{\alpha}^\beta A_\beta)_{,i})P_k=0.
$$
This is the case if and only if the potential of the electromagnetic field satisfies the system of equations:
\begin{equation}\label{5}
(\xi_{\alpha}^j A_j)_{,i} = \xi^{j}_\alpha F_{ij}, \quad F_{ji}=A_{i,j}-A_{j,i}.
\end{equation}
Unlike the free Hamilton-Jacobi equation, the equation \eqref{1} in a space with a group of motions in the general case has no integrals of motion. The system of equations \eqref{5} defines the set of admissible electromagnetic fields in which equation \eqref{1} has the first-order $r$ integrals of motion, given by the algebra of the group $G_3.$ It can be shown \cite{OVV3}, that since the vector fields  $\xi_{\alpha}^j $ define the movement group of the space  $V_4$ the system of equations \eqref{5} can be represented in the form:
   \begin{equation}\label{6}
  \mathbf{A}_{\alpha|\beta}= C^\gamma_{\beta\alpha}{\mathbf{A_{\gamma}}}, \end{equation}
  \begin{equation}\label{7}
    A_{0|\alpha} = -\xi^i_{\alpha,0}A_i,
  \end{equation}
where:
\quad $\mathbf{A}_{\alpha}=\xi_\alpha^i A_{i},\quad  A_{\alpha}=\lambda_\alpha^\beta\mathbf{A}_\beta, \quad X_{|\alpha}=\xi_\alpha^i X_{,i}, \quad \lambda_\alpha^\beta \xi^\gamma_{\beta}=\delta^\gamma_\beta,
\quad C^\gamma_{\alpha\beta} $\quad  - structural constants of the group $G_r$. For arbitrary $r$ and $n$ the following statement is true \cite{OVV3}:

If the group of motions  $G_r$  of the space  $V_n$ acts transitively on the subspace  $V_r,$   the equations \eqref{6}, \eqref{7} form a completely integrable system. This system specifies necessary and sufficient conditions for the existence of symmetry operators that are linear in momenta.

\quad

\subsection{Notations and necessary information from Petrov group classification}

Petrov classification of spacetime manifolds $V_4,$ according to the groups of motions  $G_r$ is based on the works of Petrov, Fubini, Kruchkovich (see \cite{Petrov}). The method of constructing the classification consists of using the group structural constants to find the Killing vector fields components in the simplest (canonical) holonomic coordinate system. Then, the integration of the Killing equations allows determining the components of the metric tensor. Structural constant groups  $G_3$ are known due to the classification of real groups of motions by real non-isomorphic structures for two- and three-parameter Bianchi groups \cite{Bianchi}.

According to Bianchi classification, there are nine nonisomorphic structures for three-parameter movement groups $G_3$.

Seven classes consist of solvable groups (containing a two-parameter subgroup $G_2$).

   \begin{equation}\label{20}
   \left\{\begin{array}{ll}
   G_3( I ):\qquad C^\gamma_{\alpha\beta}=0 ;\quad \cr
   G_3(II):\quad C^\alpha_{12}=0,\quad C^\alpha_{13}=0 \quad C^\alpha_{23}= \delta^\alpha _1;\quad \cr
   G_3(III):\qquad C^\alpha_{12}=0, \quad C^\alpha_{13}=\delta^\alpha_1 \quad C^\alpha_{23}=0;\quad \cr
   G_3(IV):\qquad C^\alpha_{12}=0, \quad C^\alpha_{13}=\delta^\alpha _1\quad C^\alpha_{23}=
   \delta^\alpha _1 + \delta^\alpha_2;\quad \cr
   G_3(V):\qquad C^\alpha_{12}=0, \quad C^\alpha_{13}=\delta^\alpha_1\quad C^\alpha_{23}=
   \delta^\alpha _2;\quad \cr
   G_3(VI):\qquad C^\alpha_{12}=0, \quad C^\alpha_{13}=\delta^\alpha_1,\quad C^\alpha_{23}=
   q\delta^\alpha_2.\quad (q\ne 0, 1);\quad \cr
   G_3(VII):\quad C^\alpha_{12}=0, \quad C^\alpha_{13}=\delta^\alpha_1\quad C^\alpha_{23}=
   2\delta^\alpha_2 \cos{\alpha},\quad \alpha=const.
   \\\end{array}\ \right.\end{equation}

Two classes consist of unsolvable groups:
\begin{equation}\label{21}
\left\{\begin{array}{ll}
G_3(VIII):\quad C^\alpha_{12}=\delta^\alpha_1, \quad C^\alpha_{13}=2\delta^\alpha_2\quad C^\alpha_{23}=    -\delta^\alpha_3.\cr
G_3(IX):\quad C^\alpha_{12}=\delta^\alpha_3, \quad C^\alpha_{13}=-\delta^\alpha_2\quad C^\alpha_{23}=\delta^\alpha_1.
 \\\end{array}\ \right.
\end{equation}
The crucial step in constructing the Petrov classification is to find the canonical coordinate system. Since, in our case, $G_3$ acts transitively on the isotropic (null) hypersurface $V^*_3$, the canonical coordinate system can be chosen as semi-geodesic. In this case, the hypersurface itself will be given by the equation:
$$
u^0=const.
$$
Every group $G_3(N)$  except  $G_3(IX)$ has a two-parameter subgroup $G_2$ Thus, we can first construct the operators of the subgroup and then define the canonical coordinate system. If the subgroup  $G_2$ is Abelian and contains a singular operator, it acts on the null subspace  $V_2^*$ of the hypersurface $V^*_3$. In this case, the canonical coordinat system can be chosen so that the operators of the group $X_1, X_2$  have the form:
$$
X_1=p_1, \quad X_2=p_2.
$$
If  $G_2$ does not contain a special operator, the operators of the group $X_1, X_2$ in the canonical coordinate system can be reduced to one of the following forms:
\begin{equation}\label{22}
   \left\{\begin{array}{ll}
   A:\qquad X_1=p_2, \quad X_2=p_3;\quad \cr
   B:\qquad X_1=p_2, \quad X_2=p_1 + u^3p_2;\quad \cr
   C:\qquad X_1=p_2,\quad X_2=p_2 + u^0p_2.\quad \cr \\\end{array}\ \right.\end{equation}

Subgroups  $G_2$  in this case, are denoted as  $G_2(K)$, where  $K$ can take the value  $ A, B, C.$ Among the unsolvable groups, only the group $G_2(VIII)$ contains an (Abelian) subgroup $G_2$. By choosing the canonical coordinate system, the operators $X_1, X_2$ can be reduced to the form:

$$
X_1=p_2,\quad X_2=p_1 + u^2p_2.
$$
In all cases, the operator $X_3$ is found from the equations of the structure, whereupon the Killing equations are integrated.

The group $G_3(IX)$  has no subgroup  $G_2$   and no special operator. In this case, the operator $X_1$  can be reduced to the form: \quad  $X_1=p_1$.\quad The remaining operators of the group and the canonical coordinate system follow from the equations of the structure.

Note that in all cases, the metric tensor components contain specific functions of the variables of the local coordinate system $[u^\alpha]$ of the hypersurface  $V^*_3$  (we call them ignored) and arbitrary functions of the variable  $u^0$. We call this variable non-ignored. As before, we will stick to the notations accepted in the work of A.Z. Petrov \cite{Petrov} with minor exceptions. For example, a non-ignored variable $x^4$ would be denoted by $u^0$ etc. The letters \quad $a, b, \alpha, \beta, \gamma $ \quad  (with and without indices) denote functions that depend only on the variable $u^0)$.

\quad

\section{Solvable \boldmath{$G_3$} groups. Killing vector fields do not depend on a non-ignored variable}

\quad

When a three-parameter group of motions acts transitively on a null hypersurface, the components of the vector $ \xi^i_\alpha $  may depend on the non ignored variable  $ u^0 $. In this section, we consider groups in which \quad $(\xi^\alpha_\beta)_{,0} = 0.$ \quad According to \eqref{7} this implies: \quad $ A_0 = A_0 (u^0) $, \quad  which is equivalent to the condition:
$$
A_0=0.
$$
This gauge of potentials is used in Sections  $3-4$. Each subsection is devoted to integration of joint system \eqref{6} for specific groups. The metrics and group operators are given in the canonical coordinate
system (taken from \cite{Petrov})). In addition, the explicit form of system (2.7) and its solutions are given, as well as holonomic
components of the vector potential $A_i$, which are calculated in accordance with the given relations \eqref{6}, \eqref{7}. Note another fact that distinguishes the variant considered in this paper from the case with homogeneous spaces.  For all $ G_3$ groups (except $ G_3(IX)$)  acting on isotropic hypersurfaces, there are several nonequivalent sets of Killing vectors, depending on which $ G_2(K)$ subgroup they contain (see \eqref{22}). Therefore, for each such set, there are several nonequivalent solutions of the Killing equations. Groups $ G_3(N)$, with a subgroup $ G_2(K)$, will be denoted  $G_3(N[K])$.

The following are the results of the system of equations \eqref{6} \eqref{7} integration in the following order. First, using information from  work \cite{Petrov}, the metrics of the spaces on which the considered groups, group operators, and structure constants act are presented. Then the matrix \quad $\hat{\lambda}$,\quad the nonholonomic and the holonomic components of the vector potential of the electromagnetic field  ($\mathbf{A}_\alpha$ and $A_\alpha$) are given (with explanations of the integration procedure, if necessary).

\quad

{\bf{\subsection{ Groups \boldmath $G_3(II) - G_3(VI)$ with the singular operators}}}

\quad

If the groups $ G_3 (II) - G_3 (VI) $  act on the hypersurface $ V^*_3,$  the subgroup $ G_2 $ may contain a singular operator. In this case, the subgroup  $ G_2 $ acts on the null subspace  $ V^*_2 $  of the hypersurface  $ V^*_3 $.
The metrics of appropriate spaces  and the group operators  can be represented as:
$$
ds^2=2\exp{(-ku^3)}du^0(du^1 -\varepsilon u^3 du^2) + a_1\exp{(-2lu^3)}{du^2}^2 + 2a_2\exp{(-lu^3)}du^2 du^3 +a_3{du^2}^2.
$$
Let us present operators of the group:
$$
X_1 =p_1,\quad X_2 = p_2,\quad X_3 =(ku^1 +\varepsilon u^2) p_1+lu^2p_2+p_3.
$$
and structural constants:
$$
C^\gamma_{12}=0,\quad C^\alpha_{13}=k\delta^\alpha_1, \quad C^\gamma_{23} = \varepsilon\delta^\alpha_1 + l\delta^\alpha_2.
$$
Matrix \quad $\hat{\lambda}$, \quad has the form:
$$
||\lambda^\alpha_\beta||=
\begin{pmatrix}1 & 0 & 0 \\
  0 & 1 & 0 \\
    -(ku^1 + \varepsilon u^2)& -lu^2 & 1
\end{pmatrix},\quad
$$
Let us present the set of equations \eqref{6}:
$$
\mathbf{A}_{1|\beta} = \delta_{3\beta}k\mathbf{A_1} \rightarrow \mathbf{A_1}=\alpha_1(u^0)\exp{(-ku^3)};
$$
$$
\mathbf{A}_{2|\beta} = -\delta_{3\beta}(\varepsilon\mathbf{A_1} +l\mathbf{A_2});
$$
$$
\mathbf{A}_{3|\beta} = k\delta_{1\beta}\mathbf{A_1} + \delta_{2\beta}(\varepsilon\mathbf{A_1} +l\mathbf{A_2})
$$
The set has the solutions:

\quad

$\mathbf{A)}$
\quad  $l \ne k$.
$$
\mathbf{A_1}=\alpha_1\exp{(-ku^3)};
$$

$$
\mathbf{A_2}=(\frac{\varepsilon}{k-l})\alpha_1\exp{(-ku^3)}+\alpha_2\exp{(-lu^3)}.
$$

$$
\mathbf{A_3}=\alpha_3 + k(u^1+\frac{\varepsilon u^2}{k-l})\alpha_1\exp{(-ku^3)}+lu^2\alpha_2\exp{(-lu^3)}
$$
The holonomic components of the electromagnetic potential are as follows:
\begin{equation}\label{43}
A_\alpha=\mathbf{A}_\beta \lambda^\beta_\alpha,
\end{equation}
have the form:
$$  A_1=\alpha_1\exp{(-ku^3)}, \quad
A_2=(\frac{\varepsilon}{k-l})\alpha_1\exp{(-ku^3)}+\alpha_2\exp{(-lu^3)}, \quad
A_3=\alpha_3. \quad
$$

\quad

$\mathbf{B)} \quad k=l$.

\quad

The non-holonomic  components of the electromagnetic potential are as follows:

$$
\mathbf{A_1}=\alpha_1\exp{(-ku^3)};
$$

$$
\mathbf{A_2}=(\alpha_2-\varepsilon\alpha_1 u^3)\exp{(-ku^3)}.
$$

$$
\mathbf{A_3}=\alpha_3+(k\alpha_1u^1 +  u^2(\varepsilon\alpha_1 + k(\alpha_2 - \varepsilon \alpha_1 u^3))\exp{(-ku^3)}
$$
The holonomic components of the electromagnetic potential are as follows:
$$
A_1=\alpha_1\exp{(-ku^3)}, \quad A_2=(\alpha_2 - \varepsilon\alpha_1 u^3)\exp{(-ku^3)}, \quad A_3=\alpha_3. \quad
$$

 \quad

This exhausts the classification of admissible electromagnetic fields for movement groups of spacetime  with a special operator. Groups without a special operator are considered below.

\quad

{\bf{\subsection{ Groups \boldmath $G_3(III)$ }}}

\quad

The metrics of the spaces  and the group operators  can be represented as:

$$
{ds}^2=2du^0 du^1 + 2du^0(\frac{du^2b_0 + du^3(b_1-a_0u^1)}{u^3}) +
$$
$$
2du^2du^3(\frac{a_3-a_1u^1}{{u^3}^2}) + {du^2}^2(\frac{a_1}{{u^3}^2}) + {du^3}^2(\frac{a_1 {u^1}^2-2a_3u^1+a_2}{{u^3}^2}).
$$
Let us present operators of the group:
$$
X_1 =p_1 + u^3p_2,\quad X_2 =p_2,\quad X_3 =u^2p_2+ u^3p_3.
$$
and structural constants:
$$
C^\alpha_{12}=C^\alpha_{31} =0, \quad  C^\alpha_{23}=\delta^\alpha_2.
$$
Matrix \quad $\hat{\lambda}$, \quad has the form
$$
||\lambda^\alpha_\beta||=
\begin{pmatrix}1 &-u^3 &0 \\
  0 & 1  & 0\\0&
    -\frac{u^2}{u^3}&\frac{1}{u^3}.
\end{pmatrix},\quad
$$
From the set of equations \eqref{6} it follows:
$$
\mathbf{A}_{1,\beta}=0 \quad \mathbf{A}_{2|\beta} =-\delta_{3\beta}\mathbf{A}_{2},\quad \mathbf{A}_{3|\beta} =\delta_{2\beta}\mathbf{A}_2 \rightarrow $$
$$
\mathbf{A}_{1}=\alpha_1, \quad\mathbf{A}_{2}=\frac{\alpha_2}{u^3};\quad \mathbf{A}_{3}=\alpha_3+\frac{\alpha_2 u^2}{u^3};
$$
The holonomic components of the electromagnetic potential are as follows:
$$
A_1=(\alpha_1-\alpha_2), \quad A_2=\frac{\alpha_2}{u^3},\quad A_3=\frac{\alpha_3}{u^3}.
$$

\quad

{\bf{\subsection{ Groups \boldmath $G_3(IV[A])$ }}}

\quad

The metrics of the spaces  and the group operators  can be represented as:
$$
ds^2=2du^0 du^1+2du^0(b_0 du^2 +(b_1-b_0u^1)du^3))\exp{(-u^1)} + (a_1{du^2}^2 +
$$
$$
2(a_3-a_1u^1)du^2 du^3 +(a_1{u^1}^2-2a_2u^1 +a_3){du^3}^2)\exp{(-2u^1)}.
$$
Let us present operators of the group:
$$
X_1 =p_1 + (u^2 + u^3)p_2 +u^3p_3,\quad X_2 =p_2,\quad X_3 = p_3.
$$
and structural constants:
$$
C^\gamma_{12}=\delta^\alpha_2,\quad C^\gamma_{31} = \delta^\alpha_2 +\delta^\alpha_2, \quad  C^\alpha_{23}=0. $$
Matrix \quad $\hat{\lambda}$, \quad has the form:
$$
||\lambda^\alpha_\beta||=
\begin{pmatrix}1 &-(u^2 +u^3) & -u^3 \\
  0 & 1  & 0\\0&
    0&1
\end{pmatrix},\quad
$$
From the set of equations \eqref{6} it follows:
$$
\mathbf{A_{1,1}} +\mathbf{A_{2}}(u^2 + 2u^3)+u^3\mathbf{A_{3}}=0, \quad \mathbf{A_{1,2}}=\mathbf{A_{2}}, \quad \mathbf{A_{1,3}}=\mathbf{A_{2}}+\mathbf{A_{3}};
$$
$$
\mathbf{A_{2,1}} =-\mathbf{A_{1}},\quad \mathbf{A_{2,2}}=\mathbf{A_{2,3}}=0;
$$
$$
\mathbf{A_{3,1}}=-(\mathbf{A_{2}} +\mathbf{A_{3}}),\quad
\mathbf{A_{3,2}}=\mathbf{A_{3,3}}=0;
$$
$$\rightarrow \quad \mathbf{A_{1}}=\alpha_1 + (u^2 +u^3)\mathbf{A_{2}}+ u^3\mathbf{A_{3}},\quad \mathbf{A_{2}}=\alpha_2\exp(-u^1),
$$
$$
\mathbf{A_{3}}=\alpha_3 \exp(-u^1) - u^1\mathbf{A_{2}};
$$
The holonomic components of the electromagnetic potential are as follows:
$$
A_1=\alpha_1, \quad A_2=\alpha_2\exp{-u^1},\quad A_3=(\alpha_3-\alpha_2 u^1)\exp{-u^1}.
$$

\quad

{\bf{\subsection{ Groups \boldmath $G_3(IV[B])$ }}}

\quad

The metrics of the spaces  and the group operators  can be represented as:

$$
{ds}^2=2du^0 du^1\exp u^3 + 2du^0du^3a_0 +
$$
$$
2du^2du^3(a_3\exp u^3-a_2u^1\exp 2u^3) + {du^2}^2 a_1\exp2u^3 + {du^3}^2({u^1}^2 a_1 \exp 2u^3-2a_3u^1\exp u^3+a_2).
$$
Let us present operators of the group:
$$
X_1 =p_2 + u^3p_2,\quad X_2 =p_1,+u^3p_2\quad X_3 =u^1p_1+ u^2p_2 - p_3,
$$
and structural constants:
$$
C^\alpha_{12}=0, \quad C^\alpha_{31} =\delta^\alpha_1, \quad  C^\alpha_{23}=\delta^\alpha_1+\delta^\alpha_2. $$
Matrix \quad $\hat{\lambda}$, \quad has the form:
$$
||\lambda^\alpha_\beta||=
\begin{pmatrix}-u^3& 1 &0 \\
  1 & 0  & 0\\(u^2-u^1u^3)&
    u^1&-1.
\end{pmatrix},\quad
$$
From the set of equations \eqref{6} it follows:
$$
\mathbf{A_{1,2}}=\mathbf{A_{1,1}}=0, \quad \mathbf{A_{1,3}} = \mathbf{A_{1}},\quad
\mathbf{A_{2,2}} =\mathbf{A_{2,1}}=0,\quad \mathbf{A_{2,3}}=\mathbf{A_{1}} + \mathbf{A_{2}},
$$
$$
\mathbf{A_{3,2}} =\mathbf{A_{1}},\quad \mathbf{A_{3,1}}+u^3\mathbf{A_{3,2}}=\mathbf{A_{1}}+ \mathbf{A_{2}},\quad u^1\mathbf{A_{3,1}} + u^2\mathbf{A_{3,2}} -\mathbf{A_{3,3}}=0;
$$
The solution of the set has the form:
$$
\mathbf{A_{1}}=\alpha_2\exp u^3, \quad\mathbf{A_{2}}=(\alpha_1 u^3+\alpha_2)\exp u^3,\quad \mathbf{A_{3}}=-\alpha_3+(\alpha_1(u^2 +u^1)+\alpha_2 u^1)\exp u^3.
$$
The holonomic components of the electromagnetic potential are as follows:
$$
A_1=\alpha_{2}\exp u^3, \quad A_2=\alpha_{1}\exp u^3,\quad A_3=\alpha_3-\alpha_{1}u^1\exp u^3.
$$

\quad

{\bf{\subsection{Groups \boldmath $G_3(V[A])- G_3(VI[A])$}}}

\quad

The metrics of the spaces and the group operators  can be represented as:

$$
ds^2=2du^0 du^1+2du^0(b_0 du^2 +(b_1-b_0u^1)du^3))\exp{(-u^1)} + (a_1{du^2}^2 +
$$
$$
2(a_3-a_1u^1)du^2 du^3 +(a_1{u^1}^2-2a_2u^1 +a_3){du^3}^2)\exp{(-2u^1)}.
$$

$$
ds^2=2du^0du^1 + 2du^0(b_0 du^2\exp{(-u^1)}+b_1 du^3\exp{(-qu^1)}) + a_1{du^2}^2\exp{(-2u^1)} +
$$
$$
2a_3du^2 du^3\exp{(-(q+1)u^1)} +a_3{du^3}^2\exp{(2qu^1)}.
$$
Let us present operators of the group:
$$
X_1 =p_1 + u^2p_2 +qu^3p_3,\quad X_2 =p_2,\quad X_3 = p_3.
$$
and structural constants:
$$
C^\gamma_{12}=\delta^\alpha_2,\quad C^\gamma_{31} = q\delta^\alpha_3, \quad  C^\alpha_{23}=0.
$$

If \quad $q=1,$ \quad the group of motions is of type \quad $G_3(V).$\quad In opposite case it has type \quad $G_3(VI)$.\quad

Matrix \quad $\hat{\lambda}$, \quad has the form:
$$
||\lambda^\alpha_\beta||=
\begin{pmatrix}1 &-u^2 & -qu^3 \\
  0 & 1  & 0\\0&
    0&1
\end{pmatrix},\quad
$$
From the set of equations \eqref{6} it follows:
$$
\mathbf{A_{1,1}} +u^2\mathbf{A_{1,3}}+q^2u^3\mathbf{A_{1,3}}=0, \quad \mathbf{A_{1,2}}=\mathbf{A_{2}}, \quad \mathbf{A_{1,3}}=q\mathbf{A_{3}};
$$
$$
\mathbf{A_{2,1}} =-\mathbf{A_{2}},\quad \mathbf{A_{3,1}}=-q\mathbf{A_{3}},\quad \mathbf{A_{3,2}}=\mathbf{A_{3,3}}=\mathbf{A_{2,2}}=\mathbf{A_{2,3}}=0 ;
$$

$$\rightarrow \mathbf{A_{1}}=\alpha_1 +qu^3\alpha_3\exp{-qu^1} +u^2\alpha_2 \exp{-u^1}, \quad \mathbf{A_{2}}=\alpha_2\exp{-u^1},\quad \mathbf{A_{3}}=\alpha_3\exp{-qu^1}.
$$
The holonomic components of the electromagnetic potential are as follows:
$$
A_1=\alpha_1, \quad A_2=\alpha_2\exp{-u^1},\quad A_3=\alpha_3 \exp{-qu^1}.
$$

\quad

{\bf{\subsection{Group \boldmath $G_3(VII[A])$}}}

\quad

The metrics of the spaces  and the group operators  can be represented as:

$$
{ds}^2=2du^0 du^3 + 2du^0du^1(a_4\cos(u^3\sin c)+ a_0 \sin(u^3\sin c))\exp(-u^3\cos c) +
$$
$$
2du^0du^2((a_0\sin c -a_4\cos c)\cos(u^3\sin c)-(a_0\cos c+a_4\sin c)\sin(u^3\sin c))\exp(-u^3\cos c) +
$$
$$
2du^1du^2(a_1\cos c+2a_2\cos2(u^3\sin c)+2a_3 \sin2(u^3\sin c))\exp(-2u^3\cos c) +
$$
$$
{du^1}^2(2a_1+ 2(a_3\sin c +a_2\cos c)\cos2(u^3\sin c)+2(a_3\cos c-a_2\sin c)\sin2(u^3\sin c))\exp(-2u^3\cos c) -
$$

$$
{du^2}^2(2a_1-2(a_3\sin c - a_2\cos c)\cos2(u^3\sin c)+2(a_3\cos c+a_2\sin c)\sin2(u^3\sin c))\exp(-2u^3\cos c).
$$
Let us present operators of the group:
$$
X_1 =p_1,\quad X_2 = p_2,\quad X_3 =p_3  + (2u^2\cos{c} + u^1)p_2 - u^2p_1\quad
$$
and structural constants:
$$
C^\gamma_{12}=0, \quad  C^\alpha_{13} = \delta^\alpha_2, \quad  C^\alpha_{23}=-\delta^\alpha_1 +2\delta^\alpha_2\cos{c}. $$
where \quad $ c=const.$\quad
Matrix \quad $\hat{\lambda}$, \quad has the form:
$$
||\lambda^\alpha_\beta||=
\begin{pmatrix}1 &0 & 0 \\
 0 & 1  & 0\\u^2&
   -(u^1 + 2u^2 \cos{c}) &1
\end{pmatrix},\quad
$$
The set of equations \eqref{6} has the form:
\begin{equation}\label{2.1.1}
\mathbf{A_{1,1}} = \mathbf{A_{1,2}}=0, \quad \mathbf{A_{1,3}}=\mathbf{A_{2}},\quad
\mathbf{A_{2,1}} = \mathbf{A_{2,2}}=0, \quad \mathbf{A_{2,3}}=2\mathbf{A_{2}}\cos{c}-\mathbf{A_{1}}
\end{equation}
$$
\mathbf{A}_{3|\beta}=0 \rightarrow \mathbf{A_{3}}=\alpha_3;
$$
From \eqref{2.1.1} it followers:
$$
\mathbf{A_{1}}=D_1(u^3,u^0),\quad    \mathbf{A_{2}}=D_2(u^3,u^0).
$$
Let us denote:
$$
D_2 = B(u^3,u^0)\exp({u^3\cos{c}}).
$$
Then the set \eqref{2.1.1}  can be present in the form (the dot denotes the derivative with respect to $ u^3 $):
$$
\dot{D}_1=B\exp({u^3\cos{c}}), \quad \ddot{B}+B\sin^2{c}=0 \rightarrow B=\alpha_1\sin{(\alpha_2 + u^3\sin{c})} .
$$
Solution has the form:
$$
\mathbf{A_{1}}=\alpha_1\exp({u^3\cos{c}})\sin{(\alpha_2 + u^3\sin{c})}, \quad\mathbf{A_{2}}=\alpha_1\exp({u^3\cos{c}})\sin(\sin{c+\alpha_2 + u^3\sin{c})};
$$
The holonomic components of the electromagnetic potential are as follows:
$$
A_1=\mathbf{A_{1}}, \quad A_2=\mathbf{A_{2}},
$$
$$
A_3=\alpha_3-\alpha_1\exp({u^3\cos{c}})(u^1\sin{(\sin c+\alpha_2 + u^3\sin{c})}+u^2\sin{(2\sin c+\alpha_2 + u^3\sin{c})}).
$$

\quad

{\bf{\subsection{Group \boldmath $G_3(VI[B])$}}}

The metrics of the spaces  and the group operators  can be represented as:
$$
{ds}^2=2du^0 du^1{u^3}^{(1+\omega)} + 2du^0du^3\frac{a_0}{u^3} +
$$
$$
2du^2du^3(a_3 {u^3}^{(\omega-1)}-a_1 u^1 {u^3}^{2\omega}) + a_1 {u^3}^{2\omega}{du^2}^2 +{du^3}^2(a_1 {u^1}^2 {u^3}^{2\omega}-2a_3 u^1 {u^3}^{(\omega-1)}+\frac{a_2}{{u^3}^2}),
$$
where it denoted:\quad $\omega=\frac{1}{q-1},\quad q=const.$ \quad
Let us present operators of the group:
$$
X_1 =p_2 ,\quad X_2 =p_1+u^3p_2\quad X_3 =qu^1p_1+ u^2p_2 +(1-q)p_3,
$$
and structural constants:

$$
C^\alpha_{12}=0, \quad C^\alpha_{31} =\delta^\alpha_1, \quad  C^\alpha_{23}=\delta^\alpha_1+\delta^\alpha_2. $$
Matrix \quad $\hat{\lambda}$, \quad has the form:
$$
||\lambda^\alpha_\beta||=
\begin{pmatrix}-u^3& 1 &0 \\
  1 & 0  & 0\\\frac{(u^2-qu^1u^3)}{u^3(q-1)}&
    \frac{qu^1}{u^3(q-1)}&-\frac{1}{u^3(q-1)}.
\end{pmatrix},\quad
$$
The set of equations \eqref{6}
$$
\mathbf{A}_{\alpha|\beta} = C^\gamma_{\beta\alpha}\mathbf{A}_\gamma
$$
has the form:
$$
\mathbf{A_{1,2}}=\mathbf{A_{1,1}}=0, \quad (q-1)u^3 \mathbf{A_{1,3}} = \mathbf{A_{1}}; $$
$$
\mathbf{A_{2,2}} =\mathbf{A_{2,1}}=0, \quad (q-1)u^3 \mathbf{A_{2,3}}=q\mathbf{A_{2}};
$$
$$
\mathbf{A_{3,2}} =\mathbf{A_{1}},\quad \mathbf{A_{3,1}}+u^3\mathbf{A_{3,2}}=q\mathbf{A_{2}}\quad q u^1 \mathbf{A_{3,1}} + u^2 \mathbf{A_{3,2}}+(1-q)u^3 \mathbf{A_{3,3}}=0;
$$
The solution of the set has the form:
$$\mathbf{A_{1}}=\alpha_1 {u^3}^\omega,\quad \mathbf{A_{2}}={\alpha_2}{u^3}^{(\omega+1)}, \quad    \mathbf{A_{3}}=-\alpha_3+\alpha_1 u^2{u^3}^\omega +(q\alpha_2-\alpha_1)u^1 {u^3}^{\omega+1},
$$
The nonholonomic  components of the electromagnetic potential are as follows:
$$
A_\alpha=\mathbf{A}_\beta \lambda^\beta_\alpha, \rightarrow  A_1=(\alpha_2-\alpha_1){u^3}^{(\omega+1)}, \quad A_2=\alpha_1{u^3}^{\omega},\quad A_3=\frac{\alpha_3}{u^3}-\alpha_1u^1{u^3}^\omega .
$$

\quad

{\bf{\subsection{Group \boldmath $G_3(VII[B])$}}}

\quad

The metrics of the spaces  and the group operators  can be represented as:
$$
{ds}^2=2du^0 du^1 r_3S +2du^0du^2(a_0 - u^3){r_3}^{-1}S+ 2du^0du^3({u^3}-a_0)u^1{r_3}^{-1}S
$$
$$
2du^2du^3(a_3 {r_3}^{-3}S - a_1 u^1 a_1{r_3}^{-1}S^2) + {du^2}^2 a_1{r_3}^{-2}S^2  + {du^3}^2(a_1{u^1}^2{r_3}^{-2}S^2-2a_3{r_3}^{-3}S+a_2{r_3}^{-4}).
$$
Let us present operators of the group:

$$
X_1 =p_2 ,\quad X_2 =p_1+u^3p_2\quad X_3 =(u^2 + u^1(u^3 -2\cos c))p_1+ u^2u^3p_2 +{r_3}^2 p_3,
$$
and structural constants:
$$
C^\alpha_{12}=0, \quad C^\alpha_{31} =\delta^\alpha_2, \quad  C^\alpha_{23}=q\delta^\alpha_2 -\delta^\alpha_1.
$$
where \quad $r_3 = ({u^3}^2 -2u^3\cos c + 1),\quad S=\exp (-2ctgc arctg{\frac{u^3 - \cos c}{\sin c}})$ \quad $a_i=a_i(u^0), \quad c=const.$ \quad
Matrix \quad $\hat{\lambda}$, \quad has the form:
$$
||\lambda^\alpha_\beta||=
\begin{pmatrix}-u^3& 1 &0 \\
  1 & 0  & 0\\\frac{u^1u^3(2 \cos c-u^3)}{r_3}&
    -\frac{u^2 + u^1(2\cos c - u^3)}{r_3}&-\frac{1}{r_3}.
\end{pmatrix},\quad
$$
The set of equations \eqref{6}
$$
\mathbf{A}_{\alpha|\beta} = C^\gamma_{\beta\alpha}\mathbf{A_\gamma}
$$
has the form:

\begin{equation}\label{2.1.2}
\mathbf{A_{1,2}}=\mathbf{A_{1,1}}=0, \quad r_3\mathbf{A_{1,3}} + \mathbf{A_{2}}=0;
\end{equation}

\begin{equation}\label{2.1.3}
\mathbf{A_{2,2}}=\mathbf{A_{2,1}}=0, \quad r_3\mathbf{A_{2,3}} -\mathbf{A_{1}} +q\mathbf{A_{2}}=0;
\end{equation}

\begin{equation}\label{2.1.4}
\mathbf{A_{3,2}}=\mathbf{A_{2}}, \quad \mathbf{A_{3,1}} +u^3\mathbf{A_{3,2}}+\mathbf{A_{1}} -2\cos c \mathbf{A_{2}}=0,
\end{equation}
$$
\quad (u^2 + u^1(2\cos{c}-u^3))\mathbf{A_{3,1}} +u^2u^3\mathbf{A_{3,2}}+ r_3\mathbf{A_{3,3}}=0;
$$
From \eqref{2.1.2}, \eqref{2.1.3} it followes:
$$
\mathbf{A_{1}}=B(u^0,u^3),\quad \mathbf{A_{2}}=-r_3B_{,3}. \quad
$$
The function
$ B $
satisfies the equation:
\begin{equation}\label{2.1.5}
r_3(r_3B_{,3} )_3+2(r_3B_{,3})\cos{c}+B=0.
\end{equation}
The solution of the equation \eqref{2.1.5} has the form:
$$
B=\sin{(\alpha_2+arctg{(\frac{u^3-\cos{c}}{\sin{c}}})})\exp{(\alpha_1-
(ctg{c})arctg{(\frac{u^3-\cos{c}}{\sin{c}}})})
$$
The first two equations of the system \eqref {2.1.4} have a solution
$$
\mathbf{A_{3}}=u^1((u^3-2\cos{c})r_3 B_{,3}-B)-u^2 r_3 B_{,3}+b_3(u^0,u^3).
$$
From the last equation it follows: \quad $ b_3 = \alpha_3. $ \quad As a result, we obtain the nonholonomic components of the potential of the electromagnetic field:
$$
\mathbf{A_{1}}= \sin{(\alpha_1+arctg{(\frac{u^3-\cos{c}}{\sin{c}}})})\exp{(\alpha_2-
(ctg{c})arctg{(\frac{u^3-\cos{c}}{\sin{c}}})}),
$$

$$
\mathbf{A_{2}}= -\sin{(\alpha_1-c+arctg{(\frac{u^3-\cos{c}}{\sin{c}}})})\exp{(\alpha_2-
(ctg{c})arctg{(\frac{u^3-\cos{c}}{\sin{c}}})})
$$
$$
\mathbf{A_{3}}=\alpha_3  -((u^2 + u^1(2\cos{c}))\sin{(\alpha_1-c+arctg{(\frac{u^3-\cos{c}}{\sin{c}}})})+
$$
$$
u^1\sin{(\alpha_1+arctg{(\frac{u^3-\cos{c}}{\sin{c}}})}))\exp{(\alpha_2-
(tg{c})arctg{(\frac{u^3-\cos{c}}{\sin{c}}})})
$$
The holonomic components of the electromagnetic potential are as follows:
$$
A_1=-\exp{(\alpha_1-ctg{c}arctg{\frac{(u^3-\cos{c})}{\sin{c}}})}(\sin{(\alpha_1-c+arctg{\frac{(u^3-\cos{c})}{\sin{c}}}})+
$$
$$
u^3\sin{(\alpha_1 +c+arctg{\frac{(u^3-\cos{c})}{\sin{c}}}}),
$$

$$
\quad A_2=\exp{(\alpha_1-(ctg{c})arctg{\frac{(u^3-\cos{c})}{\sin{c}}})}\sin{(\alpha_0-c+arctg{\frac{(u^3-\cos{c})}{\sin{c}}})},
$$
$$
\quad A_3=\frac{\alpha_3}{(u^3-\cos{c})^2 +{\sin{c}}^2}-u^1\exp{(\alpha_1-(ctg{c})arctg{\frac{(u^3-\cos{c})}{\sin{c}}})}\sin{(\alpha_1+arctg{\frac{(u^3-\cos{c})}{\sin{c}}})}.
$$

\section{Insolvable Groups \boldmath{$G_3(N) $}}

\quad

{\bf{\subsection{ Groups \boldmath $G_3(VIII)$}}}

\quad

The metrics of the spaces  and The group operators  can be represented as:

Let us present operators of the group:
$$
X_1 =p_2 ,\quad X_2 =p_3+u^2p_2\quad X_3 =\exp{u^3}p_1+({u^2}^2 +\varepsilon\exp{{u^3}^2})p_2+2u^2p_3,
$$
where \quad $\varepsilon=0,-1,+1,$

and structural constants:
$$
C^\alpha_{12}=0, \quad C^\alpha_{13} =2\delta^\alpha_2, \quad  C^\alpha_{23}=\delta^\alpha_2 -\delta^\alpha_1. $$
Matrix \quad $\hat{\lambda}$, \quad has the form:
$$
||\lambda^\alpha_\beta||=
\begin{pmatrix}({u^2}^2-\epsilon\exp{2u^3})\exp-u^3& -2u^2\exp{-u^3} &\exp-u^3 \\
  1 & 0  & 0\\-u^2&-1
    &0.
\end{pmatrix},\quad
$$
The set of equations \eqref{6} has the form:

$$
\mathbf{A_{1,2}}=0,\quad\mathbf{A_{1,3}}=-\mathbf{A_{1}}, \quad \mathbf{A_{1,1}} = 2(u^2\mathbf{A_{1}}-\mathbf{A_{2}})\exp-u^3;
$$

$$
\mathbf{A_{2,2}}=\mathbf{A_{1}}, \quad \mathbf{A_{2,3}}=-u^2\mathbf{A_{1}},\quad \mathbf{A_{2,1}}\exp u^3+({u^2}^2-\varepsilon\exp2u^3)\mathbf{A_{1}}+\mathbf{A_{3}}=0;
$$

$$
\mathbf{A_{3,2}}=2\mathbf{A_{2}}, \quad \mathbf{A_{3,3}}=\mathbf{A_{3}}-2u^2\mathbf{A_{2}} \quad \mathbf{A_{3,1}}\exp u^3+2({u^2}^2+\varepsilon\exp2u^3)\mathbf{A_{2}}+2u^2\mathbf{A_{3,3}}=0;
$$

This implies:
$$
\mathbf{A_{1}}=2B(u^1,u^0)\exp-u^3,\quad \mathbf{A_{2}}=2B u^2\exp(-u^3)-\dot{B}, \quad    $$
$$
\mathbf{A_{3}}=(\ddot{B}-2\varepsilon B)\exp(-u^3)-2u^2\dot{B}+2B{u^2}^2 \exp{u^3}),
$$
where $ B $ is a function of  $ (u^0, u^1) $ satisfying the equation:
\begin{equation}\label{2.1.4}
\dot{B}_{,11}=4\varepsilon\dot{B} \quad \rightarrow \quad  \dot{B}_{,1}=\alpha+4\varepsilon B.
\end{equation}
The dots denote the derivatives with respect to $u^1$.
The holonomic components of the electromagnetic potential are as follows:
$$
A_1=\alpha,\quad A_2= 2B\exp(-u^3), \quad
A_3=-\dot{B}
$$
There are three different metrics. Depending on the value of $ \varepsilon $, the function  $ B $ and the metric of the space admitting this group have the form

\quad

1. $\varepsilon=0$:
$$
B=c {u^1}^2+\beta u^1+\gamma;
$$
$$
{ds}^2=2du^0 du^1 + 2du^0du^2(a_0 - 2u^1 a_4 -{u^1}^2)\exp-u^3 +2du^0du^3(a_4 - u^1) +
$$
$$
2du^2du^3(2a_1 u^1 + a_2)\exp-u^3 + {du^2}^2(4a_1{u^1}^2 + 4a_2 u^1 + a_3)\exp-2u^3 + {du^3}^2a_1.
$$

2. $\varepsilon=-1$:

$$
B=\alpha_1\sin2(u^1+\alpha_2)+\frac{\alpha}{4};
$$
$$
{ds}^2=2du^0 du^1 + 2du^0du^2(a_4\cos2u^1+a_0 \sin2u^1
-\frac{1}{2})\exp-u^3 +2du^0du^3(a_0\cos 2u^1-a_4 \sin 2u^1
) +
$$
$$
2du^2du^3(a_3\cos 4u^1 - a_2 \sin 4u^1)\exp-u^3 + {du^2}^2(a_2\cos 4u^1 + a_3 \sin 4u^1
-\frac{a_1}{2})\exp-2u^3 -
$$
$$
{du^3}^2(a_2\cos 4u^1+a_3 \sin 4u^1
+\frac{a_1}{2}).
$$
\quad

3. $\varepsilon=1               $:

$$
B=\alpha_1sh2u^1+\alpha_2sh2u^1-\frac{\alpha}{4}.
$$

$$
{ds}^2=2du^0 du^1 + 2du^0du^2(a_4\exp -2u^1+ a_0\exp 2u^1
+\frac{1}{2})\exp-u^3 +2du^0du^3(a_4\exp-2u^1- a_0 \exp2u^1
) +
$$
$$
2du^2du^3(a_2\exp-4u^1+ a_3\sin 4u^1)\exp-u^3 + {du^2}^2(a_2\exp-4u^1 + a_3 \exp 4u^1
+\frac{a_1}{2})\exp-2u^3 +
$$
$$
{du^3}^2(a_2\exp-4u^1 + a_3 \exp 4u^1 -\frac{a_1}{2}).
$$

\quad

{\bf{\subsection{Groups \boldmath $G_3(IX)$}}}

\quad

The metrics of the spaces  and the group operators  can be represented as:
$$
{ds}^2=2du^0du^1 + {du^2}^2(a_1\sin 2u^1 - a_2\cos 2u^1 +a_3){\cos}^2 u^3 -
$$
$$
2du^2du^3(a_1\cos 2u^1 + a_2\sin 2u^1)\cos u^3 +  {du^3}^2(a_2\cos 2u^1 - a_1\sin 2u^1 +a_3)
+
$$
$$
2du^0du^2(\sin u^3+(a_0\cos u^1 + a_4\sin u^1)\cos u^3)
+ 2du^0du^3(a_0\sin u^1 - a_4 \cos u^1).
$$

Let us present operators of the group:

$$
X_1 =p_2 ,\quad X_2 =\frac{\cos u^2}{\sin u^3}p_1-tg u^3 \cos u^2 p_2+\sin u^2 p_3
$$

$$
\quad X_3=\partial_2(X_2)=-\frac{\sin u^2}{\sin u^3}p_1+tg u^3 \sin u^2 p_2+\cos u^2 p_3.
$$
and structural constants:
$$
C^\alpha_{12}=\delta^\alpha_3, \quad C^\alpha_{13} =-\delta^\alpha_2, \quad  C^\alpha_{23}=\delta^\alpha_1. $$
Matrix \quad $\hat{\lambda}$, \quad has the form:
$$
||\lambda^\alpha_\beta||=
\begin{pmatrix}\frac{{\sin u^3}^2}{\cos u^3}& \cos u^2 \sin u^3 &-\sin u^2 \sin u^3 \\
  1 & 0  & 0\\0 &\sin u^2&\cos u^2.
\end{pmatrix},\quad
$$
The set of equations \eqref{6} has the form:

$$
\mathbf{A_{1,2}}=0,\quad\mathbf{A_{1,1}}\frac{\cos u^2}{\sin u^3}+\mathbf{A_{1,3}}\sin u^2+\mathbf{A_{3}}=0,
$$
$$
\quad \mathbf{A_{1,1}}\frac{\sin u^2}{\sin u^3}-\mathbf{A_{1,3}}\cos u^2+\mathbf{A_{2}}=0;
$$

$$
\mathbf{A_{2,2}}=\mathbf{A_{3}},\quad\mathbf{A_{2,1}}\frac{\cos u^2}{\sin u^3}-\mathbf{A_{2,2}}tg u^3 \cos u^2+\mathbf{A_{2,3}}\sin u^2=0,
$$
$$
-\mathbf{A_{2,1}}\frac{\sin u^2}{\sin u^3}+ \mathbf{A_{2,2}}tg u^3 \sin u^2+\cos u^2\mathbf{A_{2,3}}+\mathbf{A_{1}}=0;
$$

$$
\mathbf{A_{3,2}}=-\mathbf{A_{2}},\quad\mathbf{A_{3,1}}\frac{\cos u^2}{\sin u^3}- \mathbf{A_{3,2}}tg u^3 \cos u^2+\mathbf{A_{3,3}}\sin u^2-\mathbf{A_{1}}=0,
$$
$$
-\mathbf{A_{3,1}}\frac{\sin u^2}{\sin u^3}+ \mathbf{A_{3,2}}tg u^3 \sin u^2+\cos u^2\mathbf{A_{3,3}}=0.
$$

This implies:
$$
\mathbf{A_{1}}=\alpha_1\sin u^3, \quad \mathbf{A_{2}}=\alpha_1 \cos u^2\cos u^3, \quad
\mathbf{A_{3}}=-\alpha_1 \cos u^3\sin u^2.
$$
The holonomic components of the electromagnetic potential are as follows:
$$
A_1=\alpha_1 tg u^3,\quad A_2=\alpha_1 \cos u^2 \cos u^3, \quad
A_3=0.
$$

\section{Killing vector fields depend on the non-ignored variable $u^0$}

\quad

{\bf{\subsection{Group \boldmath $G_3(II[C])$}}}

\quad

The metrics of the spaces  and the group operators  can be represented as:
$$
ds^2=2{du^0}(a_0 du^1 + \varepsilon(2{a_2}u^1+a_3) du^2) + (a_1 + \varepsilon(2{a_2}{u^1}^2 + 3a_3 u^1 +a_4)u^1)du^3) +
$$
$$
\varepsilon u^1(a_2 {u^1}^3 + 2a_3 {u^1}^2 +a_4u^1 +2a_1){du^0}^2 + 4(a_3 + 2a_2u^1)du^3 du^2 +4a_2{du^2}^2 +(a_4+4a_3 u^1 +  4a_2{u^1}^2) {du^3}^2.
$$
Let us present operators of the group:
$$
\quad X_1 =p_2,\quad X_2 = p_3,\quad X_3 =-p_1 +u^3p_2 +\varepsilon u^0 p_3,
\quad \varepsilon=0,1.
$$
and structural constants:
$$
C^\gamma_{12}=C^\gamma_{13} = 0,\quad  C^\alpha_{23}=\delta^\alpha_1. $$
Matrix \quad $\hat{\lambda}$, \quad has the form:
$$
||\lambda^\alpha_\beta||=
\begin{pmatrix}u^3 &\varepsilon u^0 & -1 \\
  1 & 0  & 0\\0&
    1&0
\end{pmatrix},\quad
$$
From the set of equations \eqref{6}
$$
\mathbf{A_{\alpha|\beta}} = C^\gamma_{\beta\alpha}\mathbf{A_\gamma}
$$
it follows:
$$
\mathbf{A_{1|\beta}} = 0 \quad \rightarrow \mathbf{A_{1,\beta}}=0 \rightarrow  \mathbf{A_{1}}=2\alpha_1;
$$
$$
\mathbf{A_{2|\beta}} =-\delta_{3\beta}\mathbf{A_{1}}, \rightarrow \mathbf{A_{2}}=2\alpha_1u^1+\alpha_2;
$$
$$
\mathbf{A_{3|\beta}} =\delta_{2\beta}\mathbf{A_1}, \rightarrow  \mathbf{A_{3,2}}=0, \quad \mathbf{A_{3,3}}=\mathbf{A_{1}}, \quad \mathbf{A_{3,1}}=\varepsilon u^0\mathbf{A_{1}}
\rightarrow \mathbf{A_{3}}=2\alpha_1(u^3 + \varepsilon u^0 u^1)-\alpha_3;
$$
The holonomic components of the electromagnetic potential are as follows:
$$
A_\alpha=\mathbf{A}_\beta \lambda^\beta_\alpha, \rightarrow  A_1=\alpha_3+\alpha_2\varepsilon u^0, \quad A_2=2\alpha_1,\quad A_3=\alpha_2+2\alpha_1 u^1.
$$
The holonomic component\quad $A_0$\quad can be found from the equation:
$$
A_{0|\alpha}=-\xi^\beta_{\alpha,0}A_\beta, \rightarrow A_0=\varepsilon(\alpha_1 u^1+\alpha_2)u^1 + \alpha_0.
$$

\quad

{\bf{\subsection{Group \boldmath $G_3(III[C])$}}}

\quad

The metrics of the spaces  and the group operators  can be represented as:
$$
ds^2=(a_0+\varepsilon a_2 {u^1}^2){du^0}^2 + 2(a_1du^1- \varepsilon u^1(a_3\exp{(-u^1)}du^3 + a_2 du^2))du^0 + a_2{du^2}^2 +
$$
$$
2a_3\exp{(-u^1)}du^2du^3 +a_4\exp{(-2u^1)}{du^3}^2.
$$
Let us present operators of the group:
$$
X_1 =p_1 +\varepsilon u^0 p_2+u^3p_3, \quad X_2 =p_2,\quad X_3 = p_3,\quad \varepsilon=0,1.
$$
and structural constants:
$$
C^\gamma_{12}=0,\quad C^\alpha_{13}=-\delta^\alpha_3, \quad C^\gamma_{23} = 0. $$
Matrix \quad $\hat{\lambda}$, \quad has the form:
$$
||\lambda^\alpha_\beta||=
\begin{pmatrix}1 & -\varepsilon u^0 & -u^3 \\
  0 & 1 & 0 \\
    0& 0 & 1
\end{pmatrix},\quad
$$
The set of equations \eqref{6} has the form:
$$
\mathbf{A_{1|\beta}} = \delta_{3\beta}\mathbf{A_3} \rightarrow \mathbf{A_{1,3}}=\mathbf{A_{3}};
$$
$$
\mathbf{A_{2|\beta}} =0, \rightarrow \mathbf{A_{2}}=\alpha_2(u^0);
$$
$$
\mathbf{A_{3|\beta}} =-\delta_{1\beta}\mathbf{A_3}, \rightarrow  \mathbf{A_{3,1}} =-\mathbf{A_{3}}
\rightarrow \mathbf{A_{3}}=\alpha_3(u^0)\exp{(-u^1)}\rightarrow
\mathbf{A_1}=\alpha_1+u^3 \alpha_3\exp{(-u^1)};
$$
The holonomic components of the electromagnetic potential are as follows:
$$
A_1=\alpha_1-\alpha_2\varepsilon u^0, \quad A_2=\alpha_2,\quad A_3=-\varepsilon\alpha_2 u^1.
$$
The holonomic component $A_0$ can be found from the equation:
$$
A_{0|\alpha}=-\xi^\beta_{\alpha,0}A_\beta, \rightarrow A_0=\varepsilon(u^0\alpha_2 - \alpha_1)u^1.
$$

\quad

{\bf{\subsection{ Groups \boldmath $G_3(V[C])$ with the singular operators}}}

\quad

The metrics of the spaces  and the group operators  can be represented as:
$$
{ds^2}={du^0}^2 a_1 {u^1}^2\exp{2u^3}+2du^0[du^1 \exp{u^3}-du^2a_1 u^1\exp{2u^3} +
$$
$$
(a_0 - a_2 u^1 \exp{u^3})du^3] + {du^2}^2 a_1 \exp{2u^3} +2du^2 du^3 a_2\exp{u^3}
$$
Let us present operators of the group:
$$
X_1 =p_2, \quad X_2 =p_1+ u^0 p_2,\quad X_3 =u^1p_1 + u^2 p_2 - p_3,
$$
where \quad $a_i=a_i(u^0),$ and structural constants:

$$
C^\gamma_{12}=0,\quad C^\alpha_{13}=\delta^\alpha_1, \quad C^\gamma_{23} = \delta^\alpha_2. $$
Matrix \quad $\hat{\lambda}$, \quad has the form:
$$
||\lambda^\alpha_\beta||=
\begin{pmatrix}-u^0 & 1 & 0 \\
  1 & 0 & 0 \\
  u^2-u^0u^1& u^1 & -1
\end{pmatrix},\quad
$$

From the set of equations \eqref{6}

$$
\mathbf{A_{\alpha|\beta}} = C^\gamma_{\beta\alpha}\mathbf{A_\gamma}
$$
it follows:
$$
\mathbf{A_{1|\beta}} = -\delta_{3\beta}\mathbf{A_1} \rightarrow \mathbf{A_{1,3}}=\alpha_1(u^0)\exp{u^3};
$$
$$
\mathbf{A_{2|\beta}} =-\delta_{3\beta}\mathbf{A_2}, \rightarrow \mathbf{A_{2}}=\alpha_2(u^0)\exp{u^3};
$$
$$
\mathbf{A_{3|\beta}} =\delta_{1\beta}\mathbf{A_1}+\delta_{2\beta}\mathbf{A_2}, \rightarrow  \mathbf{A_{3}} = -\alpha_3(u^0)+
(\alpha_1 u^2 + (\alpha_2 -\alpha_1 u^0)u^1)\exp{u^3}.
$$
The holonomic  components of the electromagnetic potential are as follows:
$$
A_\alpha=\mathbf{A}_\beta \lambda^\beta_\alpha, \rightarrow  A_1=(\alpha_2-\alpha_1 u^0)\exp{u^3}, \quad A_2=\alpha_1\exp{u^3},\quad A_3=\alpha_3.
$$
The holonomic component $A_0$ can be found from the equation:
$$
A_{0|\alpha}=-\xi^\beta_{\alpha,0}A_\beta \rightarrow A_0=\alpha_1\exp{u^3}.
$$

\section{Conclution}
All admissible electromagnetic fields of greatest interest to gravitational theory have been found. The metric tensor for these admissible fields contains arbitrary functions of nonignorable variable, so that considerable arbitrariness is preserved for them. This arbitrariness can be used, for example, in the search for self-consistent solutions of the gravitational field equations in the General Theory of Relativity, in the Brans-Dicke scalar-tensor theory (see, e.g., \cite{A}), or in other alternative theories of gravity. The non-ignored variable is either temporary (for homogeneous spaces) or(as in this article) isotropic (null). This is important when considering cosmological problems and when obtaining and studying models of spaces with gravitational waves. Let us mention other directions for further research in the framework of the obtained classification.

First, it is possible to consider a similar problem of admissible electromagnetic fields classification for the Dirac-Fock equation since the method of noncommutative integration is also applicable to this equation (see, e.g. \cite{ASh4}). At the same time, from the physical point of view, the construction of this classification is most justified in the framework of the already obtained classification of admissible electromagnetic fields for the Klein--Gordon--Fock equation.

Second, a complement to the classification carried out in this work will be the classification of generalized privileged coordinate systems in which the basic solutions of the Klein--Gordon--Fock equation can be found by the method of noncommutative integration.

Third, the resulting classification can be used to find the basic solutions of the Klein--Gordon--Fock equation and other quantum-mechanical equations of motion by the method of noncommutative integration. Note that this problem attracts the attention of many researchers (see, e.g. \cite{Mag3}, \cite{Mag4}).

Note that group approaches remain the most effective methods for constructing and studying realistic quantum mechanical models in linear and nonlinear physics \cite{N1},\cite{N2}.



\begin{thebibliography}{99}

\bibitem{VSh1}
Shapovalov V.N., Stackel spaces. {\em Sib. Math. J.} {\bf 1979}, {\em 20}, (1117-1130). doi: org/10.1007/BF00971844.

\bibitem{Miller} W. Miller. Symmetry And Separation Of Variables. {\em Cambridge University Press:Cambridge}, {\bf 1984}), (318 pp.) doi.org/10.1017/CBO9781107325623.

\bibitem{Bagrov1}
Bagrov V.G., Obukhov V.V. Complete separation of variables in the free  Hamilton-Jacobi equation, {\em Theor. Math. Phys.} {\bf 1993}, {\em 97}, {\em 2}, (1275-1289 pp.). doi: org/10.1007/BF01016874.

\bibitem{Benenti}
Benenti S., Separability in Riemannian Manifolds,  {\em SIGMA.} {\bf2016}, {\bf 12}, , {\em  13}, (1-21 pp.); doi.org/10.3842/SIGMA.2016.013.

\bibitem{Carter}
Carter B. New family of Einstein spaces. {\em Phys.Lett.}
{\bf 1968}, {\em A.25}, {\em 9} (399-400). doi.org/10.1016/0375-9601(68)90240-5.

\bibitem{Valero1}
Rajaratnam K., Mclenaghan R.G., and  Valero C. Orthogonal separation of the Hamilton Jacobi equation on spaces of constant curvature. {\em SIGMA}, {\bf 2016}, {/bf 12}, {\em 117}, (30 pp.); doi.org/10.3842/SIGMA.2016.117.

\bibitem{Valero2}
McLenaghan R. G., Rastelli G. and Valero C. Complete separability of the Hamilton-Jacobi equation for the charged particle orbits in a Lienard-Wiehert field
{/em J. Math. Phys.}, {\bf 61}, {/bf 2020}, (122903). doi.org/10.1063/5.0030305.

\bibitem{VVO2} V.V.Obukhov. Hamilton-Jacobi equation for a charged test particle in the Stackel space of type (2.0). {\em Symmetry}, {\bf12}, {\em 2020}, 12891291. doi: 10.3390/sym12081289.

\bibitem{VVO3}  V.V. Obukhov, Hamilton-Jacobi equation for a charged test particle in the  Stackel space of type (2.1). {\em Int. J. Geom. Meth. Mod. Phys}, {\bf2020} {\bf17}, {\em14}, {2050186}. doi: 10.1142/S0219887820501868.

\bibitem{VVO4}  V.V. Obukhov V.V. Separation of variables in Hamilton-Jacobi and       Klein-Gordon-Fock equations for a charged test particle in the stackel spaces of type (1.1). {\em Int. J. Geom. Meth. Mod. Phys}. {\bf 18}, {\bf2021}, {\em 03}, (2150036); doi:10.1142/S0219887821500365,
arXiv:2012.02548 gr-qc.

\bibitem{Osetrin1}
Osetrin K.E., Filippov A.E., Osetrin E.K.,
Models of generalized scalar-tensor gravitation theories with radiation allowing the separation of variables in the eikonal equation,
{\it Russ.Phys. J.} {\bf 2018}, {\bf 61}, {\em 8}, (1383-1391pp.)
doi: 10.1007/s11182-018-1546-8

\bibitem{Osetrin2}
Osetrin E.K., Osetrin, K.E., Filippov A.E. Stationary homogeneous models of Stackel spaces of type (2.1).
{\em Russ.Phys. J.} {\bf 2020}, {\bf 63}, {\em 3}, (410-419pp.);doi: 10.1007/s11182-020-02051-1

\bibitem{Osetrin3}
Osetrin E.K., Osetrin K.E., Filippov A.E.
Spatially homogeneous conformally stackel spaces of type (3.1). {\it Russ.Phys. J.}  {\bf 2020}, {\bf 63}, {\em 3}, (403-409pp.)
doi: 10.1007/s11182-020-02050-2

\bibitem{Osetrin4}
Osetrin K., Osetrin E. Shapovalov wave-like spacetimes,
{\it Symmetry}, {\bf 2020} {\bf 12}, {\em 8}, (1372).
doi: 10.3390/SYM12081372

\bibitem{Odintsov}
Capozziello S., De Laurentis M., Odintsov D. Hamiltonian dynamics and Noether symmetries in extended gravity cosmology.
{\em Eur.Phys.J.} {\bf 2012}, {\em C72}, 2068 (22 pp.). doi: 10.1140/epjc/s10052-012-2068-0

\bibitem{A}
Mitsopoulos A., Tsamparlis M., Leon G., Paliathanasis A. New conservation laws and exact cosmological solutions in Brans--Dicke cosmology with an extra scalar field. {\em Symmetry}. {\bf 2021}, {\bf 13}, {\em 8}, (1364). doi: 10.3390/sym13081364

\bibitem{ASh1}
Shapovalov A.V., Shirokov I.V. Noncommutative integration method of linear differential equations. {\em  Theoret. math. phys.} {\bf 1995}, {\em 104:2}, (921-934). doi.org/10.1007/BF02065973.

\bibitem{ASh2}
Shapovalov A.V., Shirokov I.V. Noncommutative integration method for linear partial differential equations. functional algebras and dimensional reduction. {\em  Theoret. math. phys.} {\bf 1996}, {\em 106:1}, (1-10). doi.org/10.4213/tmf1093.

\bibitem{Petrov} Petrov A. Z. Einstein Spaces, {\bf Oxford}, {\bf 1969}. ({\bf Russian original published by Nauka, Moscow, 1951}).

\bibitem{Mag1} Magazev A.A. Integrating Klein-Gordon-Fock equations in an extremal
electromagnetic field on Lie groups. {\em Theor.math.phys.}, {\bf 2012}, {\em 173:3}, (1654-1667). doi: 10.1007/s11232-012-0139-x, arxiv.org/abs/1406.5698;

\bibitem{Mag2} Magazev A. A., Shirokov I. V., Yu. A. Yurevich  Yu. A. Integrable magnetic geodesic flows on Lie groups,
{\em Theor.math. phys.}  {\bf 2008}, {\bf 156}, {\em 2}, (1127-1140); doi:org/10.4213/tmf6240.

\bibitem{OVV1}
Obukhov V.V. Algebra of Symmetry Operators for Klein-Gordon-Fock Equation. {\em Symmetry}. {\bf 2021}, {\em 13},  727 (15p.). https://doi.org/10.3390/sym13040727.

\bibitem{OVV2}
Obukhov V. V., Myrzakulov K. R., Guselnikova U. A. and Zhadyranova A. Elementary particle physics and field theory
Algebras of symmetry operators of the Klein--Gordon--Fock equation for groups acting transitively on two-dimensional subspaces of a spacetime manifold. {\em Izv. Vuz. Fizika}. {\bf 2021} {\em 7}, (126-131).

\bibitem{OVV3}
Obukhov V. V.  Algebras of integrals of motion for the Hamilton-Jacobi and Klein-Gordon-Fock equations in spacetime with a four-parameter groups of motions in the presence of an external electromagnetic field,  {\bf2021},  arXiv:2112.15138 [math-ph]

\bibitem{Bianchi}
Bianchi L. Lezioni sulla teoria dei gruppi continui finiti di trasformazioni. Pisa : Enrico Spoerri; {\bf 1918}.

\bibitem{ASh4} Shapovalov A. V.,  Breev A.I.
Non-Commutative Integration of the Dirac Equation in Homogeneous Spaces, {\em Symmetry},  {\bf 2020}, {\bf 12}, (N 11), (1867); doi.org/10.3390/sym12111867, arXiv: math-ph/2011.06401.

\bibitem{Mag3}
Magazev A.A. Constructing a complete integral of the Hamilton-Jacobi equation
on pseudo-riemannian spaces with simply transitive groups of motions
{\em Mathematical Physics Analysis and Geometry}. {\bf 2021}, {\bf 24(2):11}.  doi: 10.1007/s11040-021-09385-3.

\bibitem{Mag4} A.A.Magazev, M.N.Boldyreva. Schrodinger
equations in electromagnetic fields: symmetries and noncommutative integration.{\em Symmetry.} {\bf 2021}, {\em 13}, 1527.
doi.org/10.3390/sym13081527

\bibitem{N1} H. Jafari, H.Tajadodi. Nematollah Kadkhoda, Dumitru Baleanu, Fractional Subequation Method for Cahn-Hilliard and Klein-Gordon Equations. {\em Abstract and Applied Analysis.} {\bf 2013}, {\em 2013}, Article ID 587179, 5p. doi.org/10.1155/2013/587179

\bibitem{N2} H. Jafari, K.Goodarzi, M.Khorshidi et al. Lie symmetry and $\mu$-symmetry methods for nonlinear generalized Camassa-Holm equation. {\em Adv Differ Equ 2021},{\bf 2021}, 322. doi.org/10.1186/s13662-021-03471-0

\end{thebibliography}
\end{document}